\documentclass{article}

\usepackage{microtype}
\usepackage{graphicx}
\usepackage{subfigure}
\usepackage{booktabs} % for professional tables
\usepackage[textsize=tiny,textwidth=.75in,color=green]{todonotes}
\usepackage{sidecap}

\newcommand{\mc}{\mathcal}
\newcommand{\mb}{\mathbb}
\newcommand{\cut}[1]{}
\newcommand{\beginsupplement}{%
        \setcounter{table}{0}
        \renewcommand{\thetable}{S\arabic{table}}%
        \setcounter{figure}{0}
        \renewcommand{\thefigure}{S\arabic{figure}}%
     }
\usepackage{hyperref}
\usepackage{multirow,array}
\usepackage{enumerate, amsmath, amsthm, amssymb}
\usepackage[accepted]{icml2019}

\icmltitlerunning{Feudal Multi-Agent Hierarchies}

\begin{document}

\twocolumn[
\icmltitle{Feudal Multi-Agent Hierarchies \\ for Cooperative Reinforcement Learning}




\begin{icmlauthorlist}
 \icmlauthor{Sanjeevan Ahilan}{sanj}
 \icmlauthor{Peter Dayan}{peter}
\end{icmlauthorlist}
\icmlaffiliation{sanj}{Gatsby Computational Neuroscience Unit, University College London, London, United Kingdom}
\icmlaffiliation{peter}{Max Planck Institute for Biological Cybernetics, 72076 T\"{u}bingen, Germany}

\icmlcorrespondingauthor{Sanjeevan Ahilan}{ahilan@gatsby.ucl.ac.uk}

\icmlkeywords{multi-agent, reinforcement learning, feudal, hierarchies}

\vskip 0.3in
]



\printAffiliationsAndNotice{}  % leave blank if no need to mention equal contribution

\begin{abstract}
We investigate how reinforcement learning agents can learn to cooperate. Drawing inspiration from human societies, in which successful coordination of many individuals is often facilitated by hierarchical organisation, we introduce Feudal Multi-agent Hierarchies (FMH). In this framework, a `manager' agent, which is tasked with maximising the environmentally-determined reward function, learns to communicate subgoals to multiple, simultaneously-operating, `worker' agents. Workers, which are rewarded for achieving managerial subgoals, take concurrent actions in the world. We outline the structure of FMH and demonstrate its potential for decentralised learning and control. We find that, given an adequate set of subgoals from which to choose, FMH performs, and particularly scales, substantially better than cooperative approaches that use a shared reward function.
\end{abstract}

\section{Introduction}
In cooperative multi-agent reinforcement learning, simultaneously-acting agents must learn to work together to achieve a shared set of goals. A straightforward approach is for each agent to optimise a global objective using single-agent reinforcement learning (RL) methods such as Q-Learning \cite{watkins1992q} or policy gradients \cite{williams1992simple, sutton2000policy}. Unfortunately this suffers various problems in general. 

First, from the perspective of any one agent, the environment is non-stationary. This is because as other agents learn, their policies change, creating a partially unobservable influence on the effect of the first agent's actions. This issue has recently been addressed by a variety of deep RL methods, for instance with  centralised training of ultimately decentralised policies \cite{lowe2017multi, foerster2018counterfactual}. However, this requires that the centralised critic or critics have access to the actions and observations of all agents during training, which may not always be possible.

A second problem is coordination; agents must learn how to  choose actions coherently, even in environments in which many optimal equilibria exist \cite{matignon2012independent}. Whilst particularly challenging when agents are completely independent, such problems can be made more feasible if a form of communication is allowed \cite{vlassis2007concise}. Nevertheless, it is difficult for agents to learn how to communicate relevant information effectively to solve coordination problems, with most approaches relying on further helpful properties such as a differentiable communication channel \cite{sukhbaatar2016learning, foerster2016learning} and/or a model of the world's dynamics \cite{mordatch2018emergence}.

Third, multi-agent methods scale poorly -- the effective state space grows exponentially with the number of agents. Learning a centralised value function therefore suffers  the curse of dimensionality, whilst the alternative of decentralised learning often appears inadequate for addressing non-stationarity. Optimising a global objective also becomes challenging at scale, as it becomes difficult to assign credit to each agent \cite{wolpert2002optimal, chang2004all}. 
%Could mention how decentralisation can reduce (perceived) dimensionality here

A clue for meeting this myriad of challenges may lie in the way in which human and animal societies are hierarchically structured. In particular, even in broadly cooperative groups, it is frequently the case that different individuals agree to be \emph{assigned} different objectives which they work towards for the benefit of the collective. For example, members of a company typically have different roles and responsibilities. They will likely report to managers who define their objectives, and they may in turn be able to set objectives to more junior members. At the highest level, the CEO is responsible for the company's overall performance.

Inspired by this idea, we propose an approach for the multi-agent domain, which organises many, simultaneously acting agents into a managerial hierarchy. Whilst typically all agents in a cooperative task seek to optimise a shared reward, in Feudal Multi-agent Hierarchies (FMH) we instead only expose the highest-level manager to this `task' reward. The manager must therefore learn to solve the  principal-agent problem \cite{jensen1976theory} of communicating subgoals, which define a reward function, to the worker agents under its control. Workers learn to satisfy these subgoals by taking actions in the world and/or by setting their own subgoals for workers immediately below them in the hierarchy.

FMH allows for a diversity of rewards. This can provide individual agents with a rich learning signal, but necessarily implies that interactions between agents will not in general be fully cooperative. However, the intent of our design is to achieve collective behaviours which are apparently cooperative, from the perspective of an outside observer viewing performance on the task objective. 

Our idea is a development of a single-agent method for hierarchical RL known as feudal RL \cite{dayan1993feudal,vezhnevets2017feudal}, which involves a `manager' agent defining subgoals for a `worker' agent in order to achieve temporal abstraction. Feudal RL allows for the division of tasks into a series of subtasks, but has largely been investigated with only one worker per manager acting at any one time. By introducing a feudal hierarchy with multiple agents acting simultaneously in FMH, we not only divide tasks over time but also across worker agents. Furthermore, we embrace the full multi-agent setting, in which observations are not in general shared across agents.  

% Key aspects of this work were to achieve temporal abstraction, supported by the division of a task into a series of subtasks. Approaches therefore considered version of feudal RL in which there is only one worker per manager acting at a given moment in time. In our work, we instead consider the case of multiple simultaneously acting workers, which allows for the division of tasks amongst workers, as well as temporally, 

We outline a method to implement FMH for concurrently acting, communicating agents trained in a decentralised fashion. Our approach pre-specifies appropriate positions in the hierarchy as well as a mapping from the manager's choice of communication to the workers' reward functions. We show how to facilitate learning of our deep RL agents within FMH through suitable pretraining and repeated communication to encourage temporally extended goal-setting.

We conduct a range of experiments that highlight the advantages of FMH. In particular, we show its ability to address non-stationarity during training, as managerial reward renders the behaviour of workers more predictable. We also demonstrate FMH's ability to scale to many agents and many possible goals, as well as to enable effective coordination amongst workers. It performs  substantially better than both decentralised and centralised approaches.

\section{Background}
%what about adding something on Mechanism Design?
\subsection{Markov Decision Processes}
Single-agent RL can be formalised in terms of Markov Decision Processes, which consist of sets of states $\mc{S}$ and actions $\mc{A}$, a reward function $r: \mc{S} \times \mc{A} \rightarrow \mb{R}$ and a transition function $\mc{T} : \mc{S} \times \mc{A} \rightarrow \Delta{(\mc{S})}$, where $\Delta{(\mc{S})}$ denotes the set of discrete probability distributions over $\mc{S}$. Agents act according to a stochastic policy $\pi : \mc{S} \times \mc{A} \rightarrow [0,1]$.  One popular objective is to maximise the discounted expected future reward, defined as $\mb{E}_{\pi}[\sum_{t=0}^{\infty} \gamma^t r(s_t, a_t)]$ where the expectation is over the sequence of states and actions which result from policy $\pi$, starting from an initial state distribution $\rho_0: \mc{S} \rightarrow [0,1]$. Here $\gamma \in [0,1)$ is a discount factor and $t$ is the time step. This objective can be equivalently expressed as $\mb{E}_{s \sim \rho^\pi, a \sim \pi}[r(s,a)]$, where $\rho^\pi$ is the discounted state distribution induced by policy $\pi$ starting from $\rho_0$.

\subsection{Deterministic Policy Gradient Algorithms}
Deterministic policy gradients (DPG) is a frequently used single-agent algorithm for continuous control using model-free RL \cite{silver2014deterministic}. It uses deterministic policies $\mu_\theta : \mc{S} \rightarrow \mc{A}$, whose parameters $\theta$ are adjusted in an off-policy fashion using an exploratory behavioural policy to perform stochastic gradient ascent on an objective $J(\theta)= \mb{E}_{s \sim \rho^\mu, a \sim \mu_\theta}[r(s,a)]$.

We can write the gradient of $J(\theta)$ as:
\begin{equation}
\nabla_\theta J(\theta)= \mb{E}_{s \sim \rho^\mu}[ \nabla_\theta \mu_{\theta}(s) \nabla_a Q^\mu (s,a)|_{a=\mu_{\theta}(s)}].
\end{equation}
Deep Deterministic Policy Gradients (DDPG) \cite{lillicrap2015continuous} is a variant with policy $\mu$ and critic $Q^\mu$ being represented via deep neural networks. Like DQN \cite{mnih2015human}, DDPG stores experienced transitions in a replay buffer and replays them during training.

\subsection{Markov Games}
A partially observable Markov game (POMG) \cite{littman1994markov, hu1998multiagent}
for $N$ agents is defined by a set of states $\mc{S}$, and sets of actions $\mc{A}_1,...,\mc{A}_N$ and observations $\mc{O}_1,...,\mc{O}_N$ for each agent. In general, the stochastic policy of agent $i$ may depend on the set of action-observation histories $H_i \equiv (\mc{O}_i \times \mc{A}_i)^*$ such that $\pi_i : \mc{H}_i \times \mc{A}_i \rightarrow [0,1]$. In this work we restrict ourselves to history-independent stochastic policies $\pi_i : \mc{O}_i \times \mc{A}_i \rightarrow [0,1]$. The next state is generated according to the state transition function $\mc{T} : \mc{S} \times \mc{A}_1 \times ... \times \mc{A}_n \rightarrow \Delta{(\mc{S})}$. Each agent $i$ obtains deterministic rewards defined as $r_{i} : \mc{S} \times \mc{A}_1 \times ... \times \mc{A}_n \rightarrow \mb{R}$ and receives a private observation $\mathbf{o}_i : \mc{S} \rightarrow \mc{O}_i$. There is an initial state distribution $\rho_0 : \mc{S} \rightarrow [0,1]$ and each agent aims to maximise its own discounted sum of future rewards. 

\subsection{Centralised and Decentralised Training}
In multi-agent RL, agents can be trained in a centralised or decentralised fashion. In decentralised training, agents have access only to local information: their own action and observation histories during training \cite{tan1993multi}. Agents are often trained using single-agent methods for RL, such as Q-Learning or DDPG.

In centralised training, the action and observation histories of all agents are used, effectively reducing the multi-agent problem to a single-agent problem. Although it may appear restrictive for agents to require access to this full information, this approach has generated recent interest due to the potential for centralised training of ultimately decentralised policies \cite{lowe2017multi,foerster2018counterfactual}. For example, in simulation one can train a Q-function, known as a critic, which exploits the action and observation histories of all agents to aid the training of local policies for each one. These policies can then be deployed in multi-agent systems in the real world, where centralisation may be infeasible. 

\subsection{Multi-Agent Deep Determinstic Policy Gradients}
Multi-agent deep deterministic policy gradients (MADDPG) \cite{lowe2017multi} is an algorithm for centralised training and decentralised control of multi-agent systems. It uses deterministic polices, as in DDPG, which condition only on each agent's local observations and actions. MADDPG handles the non-stationarity associated with the simultaneous adaptation of all the agents by introducing a separate centralised critic for each agent. Lowe et al., trained history-independent feedforward networks on a range of mixed cooperative-competitive environments, significantly improving upon agents trained independently with DDPG. 

\section{Methods}

We propose FMH, a framework for multi-agent RL which addresses the three major issues outlined in the introduction: non-stationarity, scalability and coordination. In order to increase the generality of our approach, we do so without requiring a number of potentially helpful features: access to the observations of all agents (as in centralised training), a differentiable model of the world's dynamics or a differentiable communication channel between agents.

\subsection{Hierarchies}
We begin by defining the type of hierarchical structure we create for the agents. In its most straightforward setting, this hierarchy corresponds to a rooted directed tree, with the highest level manager as the root node and each worker reporting to only a single manager. Our experiments use this structure, although, for simplicity, using just two-level hierarchies. However, we also note the possibility, in more exotic circumstances, of considering more general directed acyclic graphs in which a single worker reports to multiple managers, and allowing for more than one manager at the highest level of the hierarchy. Acyclicity prevents potentially disastrous feedback cycles for the reward. 

Appropriate assignment of agents to positions in a hierarchy should depend on the varied observation and action capabilities of the agents. Agents exposed to key information associated with global task performance are likely to be suited to the role of manager, whereas agents with more narrow information, but which can act to achieve subgoals are more naturally suited to being workers. As we show in our experiments, identifying these differences can often make assigning agents straightforward. Here, we  specify positions directly, to focus on demonstrating the resulting effectiveness of the hierarchy. However, extensions in which the hierarchy is itself learned would be interesting. 
%in which managers can learn which of their workers are better suited to be sub-managers would be interesting.

\subsection{Goal-Setting}

In FMH, managers set goals for workers by defining their rewards. To arrange this, managers  communicate a special kind of message to workers that \emph{defines the workers' reward functions} according to a specified mapping. For example, if there are three objects in a room, we might allow the manager to select from three different messages, each defining a worker reward function proportional to the negative distance from a corresponding object. Our experiments explore variants of this scheme, imposing the constraint that the structure of the reward function remain fixed whilst the target is allowed to change. In this context and task, messages therefore correspond to `goals' requiring approach to different objects. The manager must learn to communicate these goals judiciously in order to solve complex tasks. In turn, the workers must learn how to act in the light of the resulting managerial reward, in addition to immediate rewards (or costs) they might also experience.

\begin{figure}[ht]
\begin{center}
\centerline{\includegraphics[width=\columnwidth]{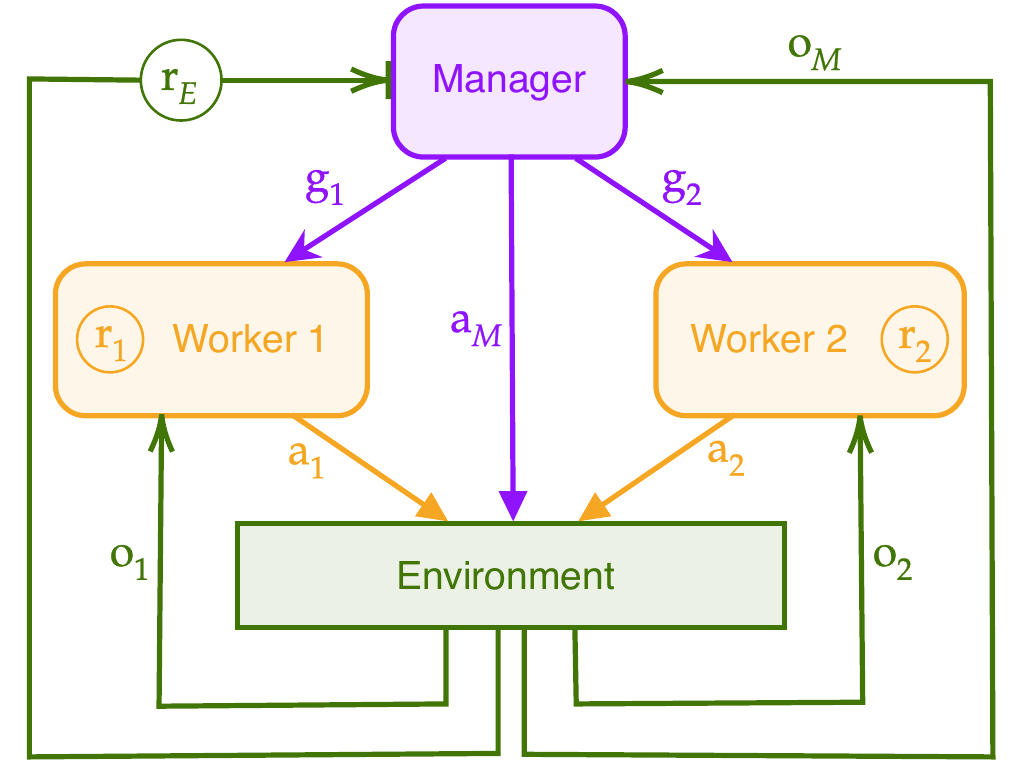}}
\caption{An example of a worker-computed Feudal Multiagent Hierarchy with one manager agent and two worker agents. Worker rewards are goal-dependent and computed locally, the manager's reward is provided by the environment.}
 \setlength\intextsep{0pt}
\label{FMH_diagram}
\end{center}
%\vskip -0.2in
\end{figure}
Since a communicated goal is simply another type of action for the manager, our approach is consistent with the formalism of a (partially-observable) Markov game, for which the reward for agent $i$ is defined as $r_{i} : \mc{S} \times \mc{A}_1 \times ... \times \mc{A}_n \rightarrow \mb{R}$. 

However, the worker's reward need not be treated as part of the environment. For instance, each worker agent $i$ could compute its own reward locally  $r_{i} : \mc{O}_i \times \mc{A}_i \rightarrow \mb{R}$, where $\mc{O}_i$ includes the observed goal from the manager.  We illustrate this `worker-computed' interpretation in Fig.~\ref{FMH_diagram}. 

% An alternative of `manager-computed' rewards is also possible, in which the manager uses its observations of the worker to determine an appropriate reward, communicating the resulting value to the worker. Unlike worker-computed rewards, manager-computed rewards require the manager to monitor performance of its workers.  

\subsection{Pretraining and Temporally Extended Subgoals}
We next consider the issue of non-stationarity. This frequently arises in multi-agent RL  because the policies of other agents may change in unpredictable ways as they learn. By contrast, in FMH we allow manager agents to determine the reward functions of workers, compelling the workers towards more predictable behaviour from the perspective of the manager. However, the same issue applies at the starting point of learning for workers: they will  not yet have learned how to satisfy the goals. We would therefore expect a manager to underestimate the value of the subgoals it selects early in training, potentially leading it  sub-optimally to discard subgoals which are harder for the worker to learn.

Thus, it would be beneficial for worker agents already to have learned to fulfill managerial subgoals. We address this issue practically in two steps. First, we introduce a bottom-up `pretraining' procedure, in which we initially train the workers before training the manager. Although the manager is not trained during this period, it still acts in the multi-agent environment and sets subgoals for the worker agents. As subgoals are initially of (approximately) equal value, the manager will explore them uniformly. If the set of possible subgoals is sufficiently compact, this will enable workers to gain experience of each potential subgoal. 
 
%  agents at the lowest level of the hierarchy before subsequently training agents one level higher \topd{this suggests that we have multiple levels -- but we don't have the space for any such expt?} and so on until finally the highest level manager is trained. SA: removed as per suggestion

This pretraining  does not completely solve the non-stationarity problem. This is because the untrained manager will, with high probability, vacillate between subgoals, preventing the workers under its command from having any reasonable hope of extracting reward. We therefore want managers not only to try out a variety of subgoals but also to \emph{commit} to them long enough that they have any hope of being at least partially achieved. Thus, the second component of the solution is a communication-repeat heuristic for the manager such that goal-setting is temporally extended. We demonstrate its effectiveness in our experiments.

\subsection{Coordination}
Multi-agent problems may have many equilibria, and good ones can only be achieved through effective coordination of agent behaviour. In FMH, the responsibility of coordination is incumbent upon the manager, which exerts control over its workers through its provision of reward. We show in the simplified setting of a matrix game, how a manager may coordinate the actions of its workers (see Suppl.\ Mat.\  \ref{coordination-game}).

\subsection{FMH-DDPG}
FMH provides a framework for rewarding agents in multi-agent domains, and can work with many different RL algorithms. In our experiments, we chose to apply FMH in combination with the single-agent algorithm DDPG,  trained in a fully decentralised fashion. As we do not experiment with combining FMH with any other algorithm, we frequently refer to FMH-DDPG simply as FMH.

\subsection{Parameter Sharing}
We apply our method to scenarios in which a large number of agents have identical properties. For convenience, when training using a decentralised algorithm (FMH-DDPG or vanilla DDPG) we implement parameter sharing among identical agents, in order to train them efficiently. We only add experience from a single agent (among those sharing parameters) into the shared replay buffer, and carry out a single set of updates. We find this gives very similar results to training without parameter sharing, in which each agent is updated using its own experiences stored in its own replay buffer (see Suppl.\ Mat.\  \ref{param-share}). 

\section{Experiments and Results}
We used the multi-agent particle environment\footnote{https://github.com/openai/multiagent-particle-envs} as a framework for conducting experiments to test the potential of our method to address the issues of non-stationarity, scalability and coordination, comparing against MADDPG and DDPG. We will release code for both the model and the environments after the reviewing process ends.

The RL agents have both an actor and a critic, each corresponding to a feedforward neural network. We give a detailed summary of all hyperparameters used in Suppl.\ Mat.\  \ref{params}. 

\subsection{Cooperative Communication}

\begin{figure*}[!ht]
%\vskip 0.2in
\begin{center}
\centerline{\includegraphics[width=\textwidth]{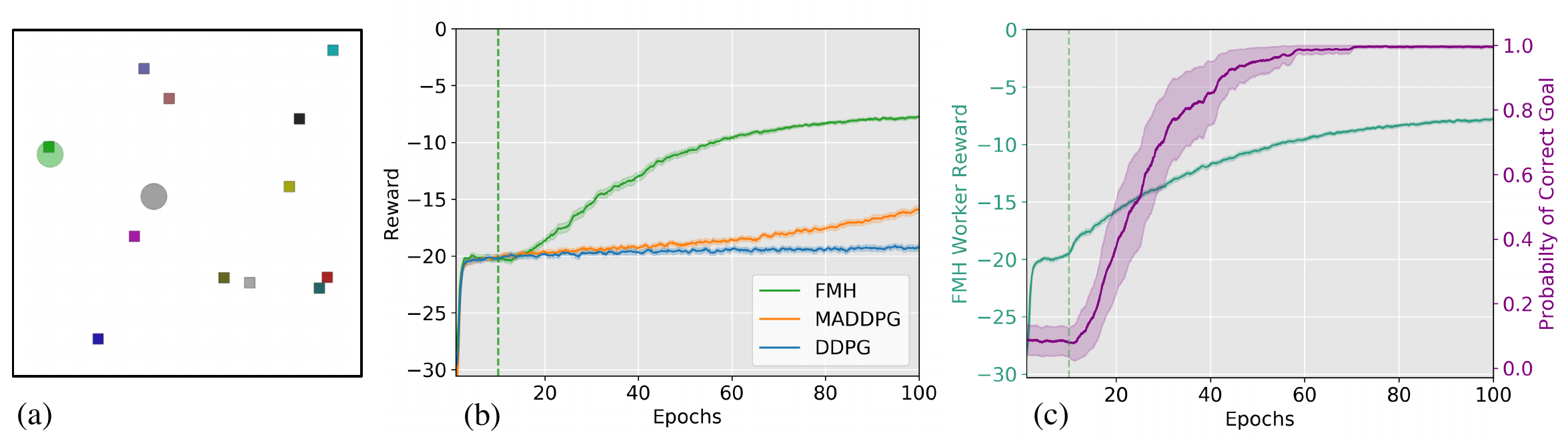}}
\caption{Cooperative Communication with 1 listener and 12 landmarks. (a) The speaker (grey circle) sees the colour of the listener (green circle) , which indicates the target landmark (green square). It communicates a message to the listener at every time step. Here there are 12 landmarks and the agent trained using FMH has successfully navigated to the correct landmark. (b) FMH substantially outperforms MADDPG and DDPG. The dotted green line indicates the end of pretraining for FMH. (c) FMH worker reward and the probability of the manager correctly assigning the correct target to the worker. The manager learns to assign the target correctly with probability 1. }
\label{speaker-listener}
\end{center}
\vskip -0.2in
\end{figure*}

We first experiment with an environment implemented in \citet{lowe2017multi} called `Cooperative Communication' (Figure~\ref{speaker-listener}a), in which there are two cooperative agents, one called the `speaker' and the other called the `listener', placed in an environment with many landmarks of differing colours. On each episode, the listener must navigate to a randomly selected landmark; and in the original problem, both listener and speaker obtain reward proportional to the negative distance\footnote{The original implementation used the negative square distance; we found this  slightly worse for all algorithms.} of the listener from the correct target. However, whilst the listener observes its relative position from each of the differently coloured landmarks, it does not know to which landmark it must navigate. Instead, the colour of the target landmark can be seen by the speaker, which cannot observe the listener and is unable to move. The speaker can however communicate to the listener at every time step, and so successful performance on the task corresponds to the speaker enabling the listener to reach the correct target. We also note that, although reward is provided during the episode, it is used only for training agents and not directly observed, which means that agents cannot simply learn to follow the gradient of reward. 

Although this task seems  simple, it is challenging for many RL methods. \citet{lowe2017multi} trained, in a decentralised fashion, a variety of single-agent algorithms, including DDPG, DQN and trust-region policy optimisation \cite{schulman2015trust} on a version of this problem with three landmarks and demonstrated that they all perform poorly on this task. Of these methods, DDPG reached the highest level of performance and so we use DDPG as the strongest commonly used baseline for the decentralised approach. We also compare our results to MADDPG, which combines DDPG with centralised training. MADDPG was found to perform strongly on Cooperative Communication with three landmarks, far exceeding the performance of DDPG.

For our method, FMH, we also utilised DDPG, but reverted to the more generalizable decentralized training that was previously ineffective. Crucially, we assigned the speaker the role of manager and the listener the role of worker. The speaker can therefore communicate messages which correspond to the subgoals of the different coloured landmarks. The listener is not therefore rewarded for going to the correct target but is instead rewarded proportional to the negative distance from the speaker-assigned target. The speaker itself is rewarded according to the original problem definition, the negative distance of the listener from the true target. 

We investigated in detail a version of Cooperative Communication with 12 possible landmarks (Figure~\ref{speaker-listener}a). FMH performed significantly better than both MADDPG and DDPG (Figure~\ref{speaker-listener}b) over a training period of 100 epochs (each epoch corresponds to 1000 episodes). For FMH, we pretrained the worker for 10 epochs and enforced extended communication over 8 time steps (each episode is 25 time steps).

In Figure~\ref{speaker-listener}c, the left axis shows the reward received by the FMH worker (listener) over training. This increased during pretraining and again immediately after pretraining concludes. Managerial learning after pretraining resulted in decreased entropy of communication over the duration of an episode (see Suppl.\ Mat.\  \ref{entropy}),  allowing the worker to optimise the managerial objective more effectively. This in turn enabled the manager to assign goals correctly, with the rise in the probability of correct assignment occurring shortly thereafter (right axis), then reaching perfection.

Our results show how FMH resolves the issue of non-stationarity. Initially, workers are able to achieve reward by learning to optimise managerial objectives, even whilst the manager itself is not competent. This learning elicits robust behaviour from the worker, conditioned on managerial communication, which makes workers more predictable from the persepective of the manager. This then enables the manager to achieve competency -- learning to assign goals so as to solve the overall task.

\begin{table*}[t]
\vskip 0.15in
\begin{center}
\begin{small}
\begin{sc}
\begin{tabular}{cc|rrrr|rrr}
%{lccccccr}
%\toprule
\hline
\multicolumn{2}{c|}{Number of} & \multicolumn{4}{c|}{Final Reward} & \multicolumn{3}{c}{Epochs until Convergence}\\
%\midrule
%\hline
Listeners & Landmarks & FMH & MADDPG & DDPG &CoM&FMH & MADDPG & DDPG\\
\hline
$1$    &$3$  &$\mathbf{-6.63 \pm 0.05}$&$\mathbf{-6.58\pm0.03}$&$-14.26\pm0.07$&$-17.28$&$56$&$24$&$55$ \\
$1$ & $6$ &$-6.91\pm0.07$&$\mathbf{-6.69\pm0.06}$&$-18.10\pm0.07$&$-18.95$&$57$&$66$&$42$ \\ 
$1$    &$12$ &$\mathbf{-7.79\pm0.06}$&$-15.96\pm0.09$&$-19.32\pm0.11$&$-19.56$ &$-$&$-$&$36$ \\
$3 $   &$6$  &$\mathbf{-7.10\pm0.04}$&$-11.13\pm0.03$&$-18.90\pm0.05$&$-18.95$&$77$&$-$&$50$ \\
$5$ & $6$  &$\mathbf{-7.17\pm0.03}$&$-18.47\pm0.04$&$-19.73\pm0.06$ &$-18.95$&$79$ &$75$ &$53$ \\
$10$    &$6$&$\mathbf{-8.96\pm0.03} $&$-19.80\pm 0.06$ &$-21.19\pm0.03$&$-18.95$&$-$&59&32 \\
\hline
\end{tabular}
\end{sc}
\end{small}
\end{center}
\vskip -0.1in
\caption{Performance of FMH, MADDPG and DDPG for versions of Cooperative Communication with different numbers of listeners and landmarks. Final reward is determined by training for 100 epochs and evaluating the mean reward per episode in the final epoch. We indicate no convergence with a $-$ symbol. For further details see Sup. Mat.~\ref{table-scaling}.}
\label{tab:agent-scaling}
\end{table*}

\begin{figure}[ht]
%\vskip 0.2in
  \begin{minipage}[c]{0.32\textwidth}%\begin{center
\centerline{\includegraphics[width=\columnwidth]{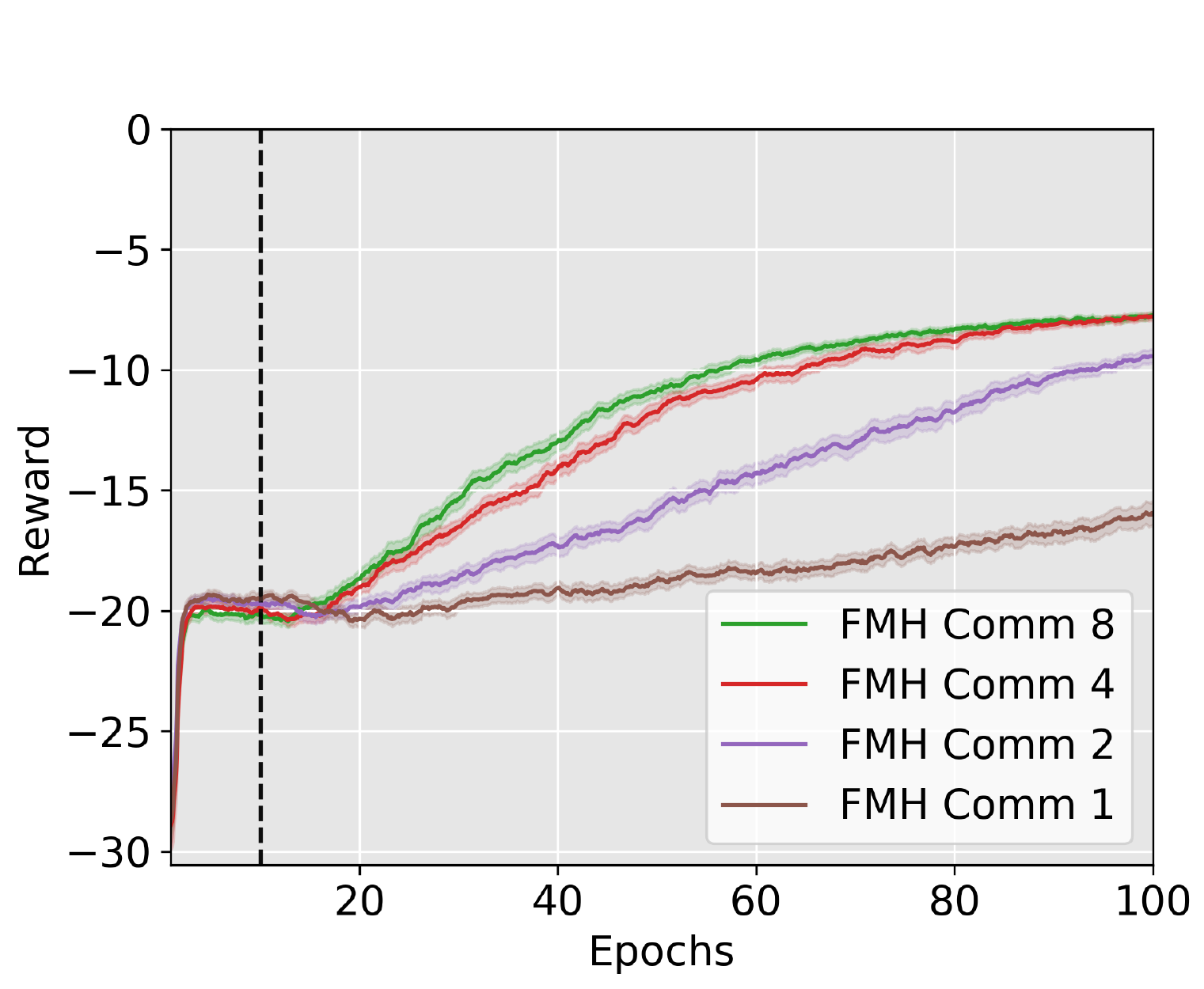}}
\end{minipage}
 \begin{minipage}[c]{0.1533\textwidth}
% \captionof{\small \textit{Figure 3}}{\small \\Communication sent by the manager is repeated for extended goal-setting over various numbers of time steps.}
\caption{\\Communication sent by the manager is repeated for extended goal-setting over various numbers of time steps.}
\label{comm-extend}%\end{center}
 \end{minipage}
%\vskip -0.2in
\end{figure}

Our implementation of FMH used both pretraining and extended goal-setting with a communication repeat heuristic. In Suppl.\ Mat.\  \ref{pretrain}, we show that pretraining the worker improved performance on this task, although even without pretraining FMH still performed better than MADDPG and DDPG. The introduction of extended communication is however more critical. In Figure~\ref{comm-extend} we show the performance of FMH with goal-setting over various number of time steps (and fixed pretraining period of 10 epochs). When there were no communication repeats (Comm 1), performance was similar to MADDPG, but introducing even a single repeat greatly improved performance. By contrast, introducing communication repeats to MADDPG did not improve performance (see Suppl.\ Mat.\ ~\ref{comm-extend-maddpg}).

\subsubsection{Scaling to many agents}

We next scaled Cooperative Communication to include many listener agents. At the beginning of an episode each listener is randomly assigned a target out of all the possible landmarks and the colour of each listener's target is observed by the speaker. The speaker communicates a single message to each listener at every time step. To allow for easy comparison with versions of Cooperative Communication with only one listener, we normalised the reward by the number of listeners. As discussed in the methods, we  shared parameters across listeners for FMH and DDPG. 

In Table~\ref{tab:agent-scaling} we show the performance of FMH, MADDPG and DDPG for variants of Cooperative Communication with different numbers of listener agents and landmarks. Consider the version with 6 landmarks, which we found that MADDPG could solve with a single listener within 100 epochs (unlike for 12 landmarks). On increasing the number of listeners up to a maximum of 10, we found that FMH scales much better than MADDPG; FMH was able to learn an effective policy with 5 listener agents whereas MADDPG could not. Further, FMH even scaled to 10 listeners, although it did not converge over the 100 epochs.

To aid interpretation of the reward values in Table~\ref{tab:agent-scaling} we also compare performance to a policy which simply moves to the centroid of the landmarks. This `Centre of Mass' (CoM) agent was trained using MADDPG until convergence on a synthetic task in which reaching the centroid is maximally rewarded, and then evaluated on the true version of the task. We find that both MADDPG and DDPG do not perform better than the CoM agent when there are 10 listeners and 6 landmarks. 

\begin{figure}[ht]
\begin{center}
\centerline{\includegraphics[width=\columnwidth]{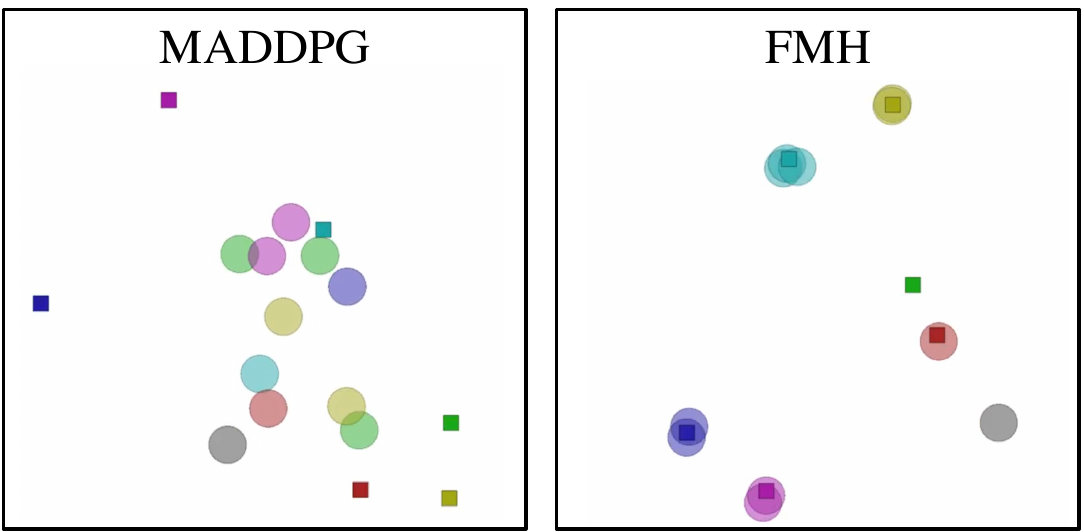}}
\caption{Scaling Cooperative Communication to 10 listeners with 6 landmarks - final time step on example episode.}
 \setlength\intextsep{0pt}
\label{10-listeners}
\end{center}
%\vskip -0.2in
\end{figure}

In Figure~\ref{10-listeners} we show the final state achieved on an example episode of this version of the task, for agents trained using MADDPG and FMH. After training for 100 epochs, MADDPG listeners do not find the correct targets by the end of the episode whereas FMH listeners manage to do so.

%will need to add column for and talk about centre of mass agent

\begin{figure*}[!ht]
%\vskip 0.2in
\begin{center}
\centerline{\includegraphics[width=\textwidth]{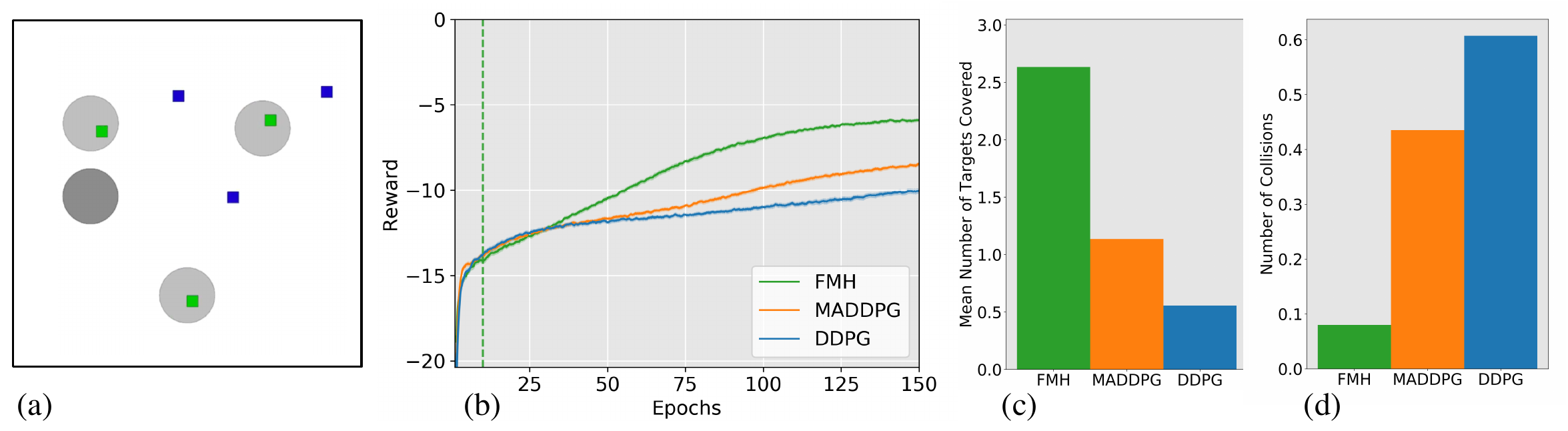}}
\caption{Cooperative Coordination (a) Three listeners (light grey agents) must move to cover the green landmarks whilst ignoring the blue landmarks. However, only the speaker (dark grey agent) can see the landmarks' colours; it communicates with the listeners at every time step. In this example, FMH agents have successfully coordinated to cover the three correct targets. (b) FMH performs significantly better than MADDPG and DDPG. The dotted green line indicates the end of pretraining for FMH. (c) Agents trained using FMH cover on average more targets, by the end of the episode, than MADDPG and DDPG (d) Agents trained using FMH avoid collisions more effectively than MADDPG and DDPG over the duration of an episode.}
\label{coop-coord}
\end{center}
\vskip -0.2in
\end{figure*}
\subsection{Cooperative Coordination}
We then designed a task to test coordination called `Cooperative Coordination' (Figure~\ref{coop-coord}a). In this task, there are 6 landmarks. At the beginning of each episode 3 landmarks are randomly selected to be green targets and 3 blue decoys. A team of 3 agents must navigate to cover the green targets whilst ignoring the blue decoys, but they are unable to see the colours of the landmarks. A fourth agent, the speaker, can see the colours of the landmarks and can send messages to the other agents (but cannot move). All agents can see each others' positions and velocities, are large in size and face penalties if they collide with each other. The task shares aspects with `Cooperative Navigation' from \cite{lowe2017multi} but is considerably more complex due to the greater number of potential targets and the hidden information. 

We apply FMH to this problem, assigning the speaker agent the role of manager. One consideration is whether the manager, the worker, or both should receive the negative penalties due to worker collisions. Here we focus on the case in which the manager only concerns itself with the `task' reward function. Penalties associated with collisions are therefore experienced only by the workers themselves, which seek to avoid these whilst still optimising the managerial objective. 

In Figure~\ref{coop-coord}b we compare the performance of FMH to MADDPG and DDPG. As with Cooperative Communication, FMH does considerably better than both after training for 150 epochs. This is further demonstrated when we evaluate the trained policies over a period of 10 epochs: Figure~\ref{coop-coord}c shows the mean number of targets covered by the final time step of the episode, for which FMH more than doubles MADDPG. Figure~\ref{coop-coord}d shows the mean number of collisions (which are negatively rewarded) during an episode. FMH collides very rarely whereas MADDPG and DDPG collide over 4 times more frequently. 

We also implement a version of Cooperative Coordination in which the manager is responsible not only for coordinating its workers but must also navigate to targets itself. We find that it can learn to do both roles effectively (see Suppl.\ Mat.\  \ref{manager-moves}).

% We find that speakers trained using MADDPG and DDPG are unable to coordinate the other agents; the speaker itself moves to cover one of the three targets but the other two agents do not move appropriately. By contrast, FMH solves this problem well. 
\subsubsection{Exploiting Diversity}
One role of a manager is to use the diversity it has available in its workers to solve tasks more effectively. We tested this in a version of Cooperative Coordination in which one of the listener agents was lighter than the other two and so could reach farther targets more quickly. 

\begin{figure}[ht]
\begin{center}
\centerline{\includegraphics[width=\columnwidth]{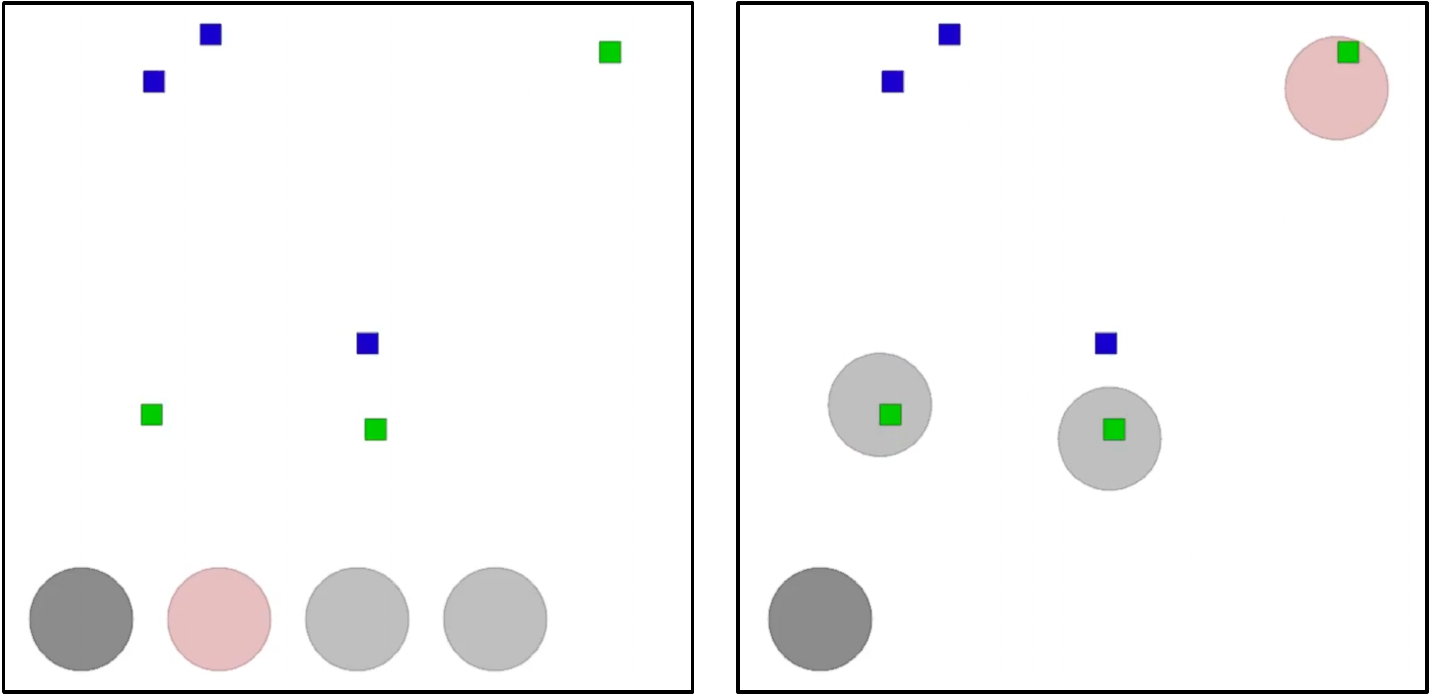}}
\caption{FMH solves the TNOF task (example episode). Left: Agents are initialised at the bottom of the environment, two targets are close by, and one far away. Right: By the end of the episode, the faster (red) agent covers the farther target on the right, despite starting on the left. }
 \setlength\intextsep{0pt}
\label{fig-diversity}
\end{center}
%\vskip -0.2in
\end{figure}

We trained FMH (without parameter sharing due to the diversity) on this task and then evaluated the trained policies on a `Two-Near, One-Far' (TNOF) version of the task in which one target is far away and the remaining two are close (see Suppl.\ Mat.\  \ref{diversity}). We did this to investigate whether FMH, which was trained on the general problem, had learned the specific strategy of assigning the farthest target to the fastest agent. We found this this to be true 100 percent of the time (evaluating over 10 epochs); we illustrate this behaviour in Figure~\ref{fig-diversity}.

\section{Discussion} 
We have shown how cooperative multi-agent problems can be solved efficiently by defining a hierarchy of agents. Our hierarchy was reward-based, with manager agents setting goals for workers,  and with the highest level manager optimising the overall task objective. Our method, called FMH, was trained in a decentralised fashion and showed considerably better scaling and coordination than both centralised and decentralised methods that used a shared reward function.

Our work was partly inspired by the feudal RL architecture (FRL) \cite{dayan1993feudal}, a single-agent method for hierarchical RL \cite{barto2003recent} which was also developed in the context of deep RL by \citet{vezhnevets2017feudal}. In particular, FMH addresses the `too many chiefs' inefficiency inherent to FRL, namely that each manager only has a single worker under its control. A much wider range of management possibilities and benefits arises when multiple agents operate at the same time to achieve one or more tasks  \cite{busoniu2008comprehensive}. We focused on the cooperative setting \cite{panait2005cooperative}; however, unlike the fully-cooperative setting, in which all agents optimise a shared reward function \cite{boutilier1996planning} our approach introduces a diversity of rewards, which can help with credit-assignment \cite{wolpert2002optimal, chang2004all} but also introduces elements of competition. This competition need not always be deleterious; for example, in some cases, an effective way of optimising the task objective might be to induce competition amongst workers, as in generative adversarial networks \cite{goodfellow2014generative}.

We followed FRL (though not all its successors; \cite{vezhnevets2017feudal} in  isolating the workers from much of the true environmental reward, making them focus on their own narrower tasks. Such reward hiding was not complete -- we considered  individualised or internalised costs from collisions that workers experienced  directly, such that their reward was not purely managerially dependent. A more complete range of possibilities for creating and decomposing rewards between managers and workers when the objectives of the two are not perfectly aligned, could usefully be studied under the aegis of  principal-agent problems  \cite{jensen1976theory, laffont2009theory}. 

Goal-setting in FMH was achieved by specifying a relationship between the chosen managerial communication and the resulting reward function. The communication was also incorporated into the observational state of the worker; however, the alternative possibility of a goal-embedding would be worth exploring \cite{vezhnevets2017feudal}. We also specified goals directly in the observational space of the workers, whereas Vezhnevets et al. specified goals in a learned hidden representation. This would likely be of particular value for problems in which defining a set of subgoals is challenging, such as those which require learning directly from pixels. More generally, work on the way that agents can learn to construct and agree upon a language for goal-setting would be most important.

To train our multi-agent systems we leveraged recent advances in the field of deep RL; in particular the algorithm DDPG \cite{lillicrap2015continuous}, which can learn in a decentralised fashion. Straightforward application of this algorithm has been shown to achieve some success in the multi-agent domain \cite{gupta2017cooperative} but also shown it to be insufficient in handling more complex multi-agent problems \cite{lowe2017multi}. By introducing a managerial hierarchy, we showed that FMH-DDPG has the potential to greatly facilitate multi-agent learning whilst still retaining the advantage of decentralised training. Our proposed approach could also be combined with centralised methods, and this would be worth further exploration. 

 Other work in multi-agent RL has also benefitted from ideas from hierarchical RL. The MAXQ algorithm \cite{dietterich2000hierarchical} has been used to train homogenous agents \cite{makar2001hierarchical}, allowing them to coordinate by communicating subtasks rather than primitive actions, an idea recently re-explored in the context of deep RL \cite{tang2018hierarchical}. A meta-controller which structures communication between agent pairs in order to achieve coordination has also been proposed \cite{kumar2017federated} (and could naturally be hybridized with FMH), as well as the use of master-slave architecture which merges the actions of a centralised master agent with those of decentralised slave agents \cite{kong2017revisiting}. Taken together, these methods represent interesting alternatives for invoking hierarchy which are unlike our primarily reward-based approach.
 
There are a number of additional directions for future work. First, the hierarchies used were simple in structure and specified in advance, based on our knowledge of the various information and action capabilities of the agents. It would be interesting to develop mechanisms for the formation of complex hierarchies. Second, in settings in which workers can acquire relevant task information, it would be worthwhile investigating how a manager might incentivise them to provide this. Third, it would be interesting to consider how a manager might learn to allocate resources, such as money, computation or communication bandwidth to enable efficient group behaviour. Finally, we did not explore how managers should train or supervise the workers beneath them, such as through reward shaping \cite{ng1999policy}. Such an approach might benefit from recurrent networks, which could enable managers to use the history of worker performance to better guide their learning.

\section{Acknowledgements}
We would like to thank Heishiro Kanagawa, Jorge A. Menendez and Danijar Hafner for helpful comments on a draft version of the manuscript. Sanjeevan Ahilan received funding from the Gatsby Computational Neuroscience Unit and the Medical Research Council. Peter Dayan received funding from the Max Planck Society. 

\bibliography{FMH_paper}
\bibliographystyle{icml2019}

% %%%%%%%%%%%%%%%%%%%%%%%%%%%%%%%%%%%%%%%%%%%%%%%%%%%%%%%%%%%%%%%%%%%%%%%%%%%%%%%
% %%%%%%%%%%%%%%%%%%%%%%%%%%%%%%%%%%%%%%%%%%%%%%%%%%%%%%%%%%%%%%%%%%%%%%%%%%%%%%%
% % DELETE THIS PART. DO NOT PLACE CONTENT AFTER THE REFERENCES!
% %%%%%%%%%%%%%%%%%%%%%%%%%%%%%%%%%%%%%%%%%%%%%%%%%%%%%%%%%%%%%%%%%%%%%%%%%%%%%%%
% %%%%%%%%%%%%%%%%%%%%%%%%%%%%%%%%%%%%%%%%%%%%%%%%%%%%%%%%%%%%%%%%%%%%%%%%%%%%%%%
% \appendix
% \section{Do \emph{not} have an appendix here}

% \textbf{\emph{Do not put content after the references.}}
% %
% Put anything that you might normally include after the references in a separate
% supplementary file.

% We recommend that you build supplementary material in a separate document.
% If you must create one PDF and cut it up, please be careful to use a tool that
% doesn't alter the margins, and that doesn't aggressively rewrite the PDF file.
% pdftk usually works fine. 

% \textbf{Please do not use Apple's preview to cut off supplementary material.} In
% previous years it has altered margins, and created headaches at the camera-ready
% stage. 
% %%%%%%%%%%%%%%%%%%%%%%%%%%%%%%%%%%%%%%%%%%%%%%%%%%%%%%%%%%%%%%%%%%%%%%%%%%%%%%%
% %%%%%%%%%%%%%%%%%%%%%%%%%%%%%%%%%%%%%%%%%%%%%%%%%%%%%%%%%%%%%%%%%%%%%%%%%%%%%%%

\clearpage
\appendix
\beginsupplement

\section{Experimental Results}
\subsection{Parameter settings}
\label{params}
In all of our experiments, we used the Adam optimizer with a learning rate of $0.001$ and $\tau = 0.01$ for updating the target networks. $\gamma$ was  0.75. The size of the replay buffer was $10^7$ and we updated the network parameters after every 100 samples added to the replay buffer. We used a batch size of 1024 episodes before making an update. For our feedforward networks we used two hidden layers with 256 neurons per layer. We trained with 10 random seeds (except otherwise stated). 

Hyperparameters were optimised using a line search centred on the experimental parameters used in \cite{lowe2017multi}. Our optimised parameters were found to be identical except for a lower value of $\gamma$ (0.75) and of the learning rate (0.001), and a larger replay buffer ($10^7$). We found these values gave the best performance for both MADDPG and FMH on a version of Cooperative Communication with 6 landmarks evaluated after 50 epochs (an epoch is defined to be 1000 episodes). 

\subsection{Parameter Sharing}
\label{param-share}
\begin{figure}[ht]
%\vskip 0.2in
  \begin{minipage}[c]{0.32\textwidth}%\begin{center}
\centerline{\includegraphics[width=\columnwidth]{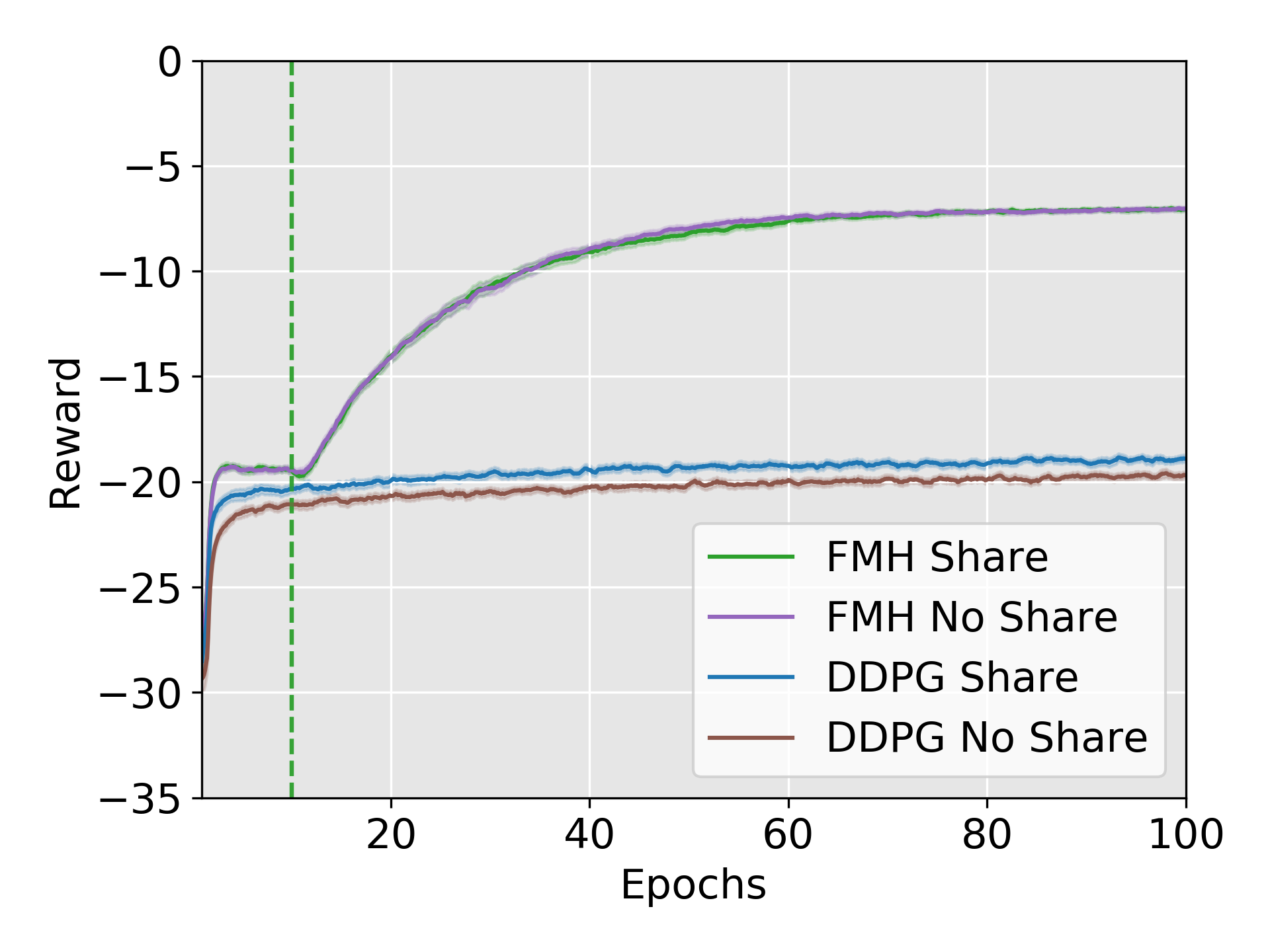}}
\end{minipage}% 
%\hfill
 \begin{minipage}[c]{0.1533\textwidth}
\caption{Parameter sharing does not affect performance for FMH but slightly improves DDPG. }
\label{share-params-fig}%\end{center}
 \end{minipage}
%\vskip -0.2in
\end{figure}

We implemented parameter sharing for the decentralised algorithms DDPG and FMH. To make training results approximately similar to implementations which do not use parameter sharing we restrict updates to a single agent and add experience only from a single agent to the shared replay buffer (amongst those sharing parameters). We find in practice that both approaches give very similar results for FMH, whereas parameter sharing slightly improves the performance of DDPG -- we show this for a version of Cooperative Communication with 3 listeners and 6 targets (Figure~\ref{share-params-fig}). In general sharing parameters reduces training time considerably, particularly as the number of agents scales. 
\subsection{Entropy of Communication}
\label{entropy}
\begin{figure}[ht]
%\vskip 0.2in
  \begin{minipage}[c]{0.32\textwidth}%\begin{center}
\centerline{\includegraphics[width=\columnwidth]{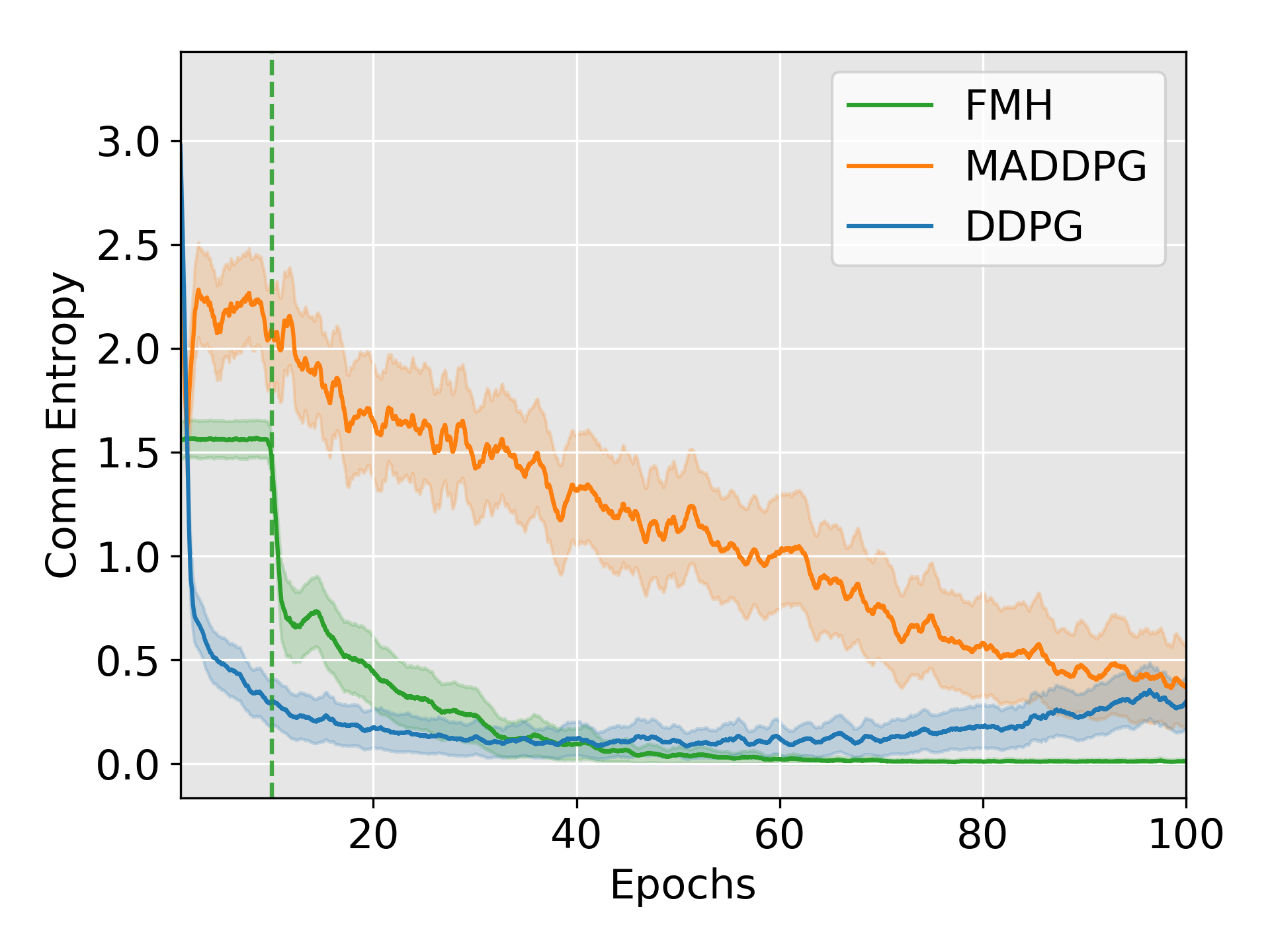}}
\end{minipage}% 
%\hfill
 \begin{minipage}[c]{0.1533\textwidth}
\caption{Entropy of managerial communication over the duration of an episode at different stages in training}
\label{entropy-fig}%\end{center}
 \end{minipage}
%\vskip -0.2in
\end{figure}

We show the change in entropy of communication over the duration of an episode for the various algorithms (Figure \ref{entropy-fig}). Agents are trained on Cooperative Communication with 12 landmarks and a single listener. As the target does not change during the middle of an episode, we expect the entropy to decrease as agents learn.

For FMH, during pretraining, entropy is high as all goals are sampled approximately uniformly (but with enforced extended communication over 8 time steps). However, shortly after pretraining ends the entropy of managerial communication rapidly decreases. This is in contrast to MADDPG which decreases in entropy more steadily.

\subsection{Pretraining} 
\label{pretrain}
Agents trained using FMH were pretrained for 10 epochs across all experiments. During pretraining all agents act in the multi-agent environment. Although all experiences are added appropriately into the replay buffers, the manager does not update its parameters during pretraining whereas the workers do.

\begin{figure}[ht]
%\vskip 0.2in
  \begin{minipage}[c]{0.32\textwidth}%\begin{center}
\centerline{\includegraphics[width=\columnwidth]{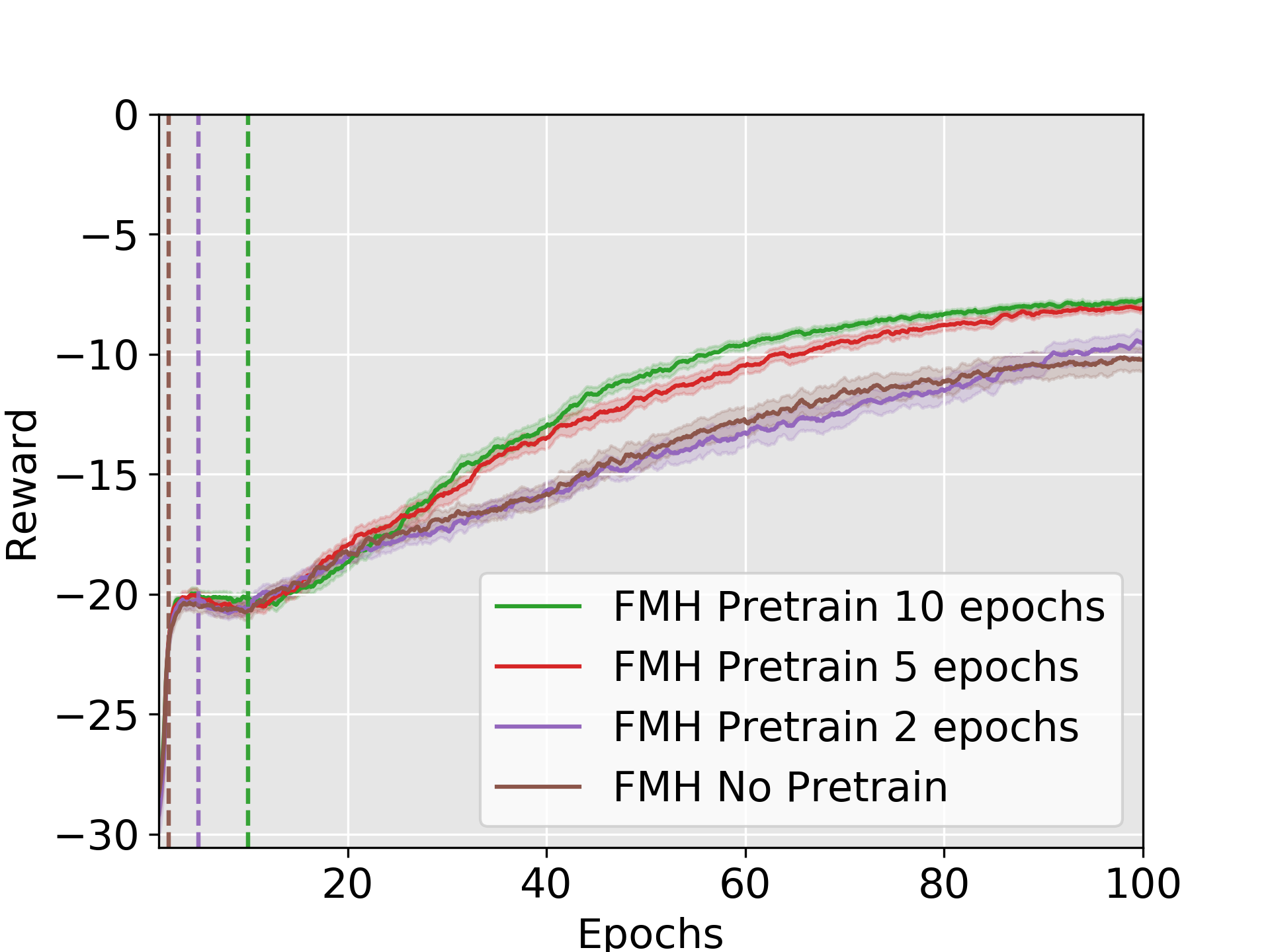}}
\end{minipage}% 
%\hfill
 \begin{minipage}[c]{0.1533\textwidth}
\caption{Pretraining improves FMH, although agents can still learn effectively without it. We use extended communication (goal-setting) for 8 time steps.}
\label{pretrain-fig}%\end{center}
 \end{minipage}
%\vskip -0.2in
\end{figure}

We show the benefits of pretraining on Cooperative Communication with 12 landmarks and a single listener agent in Figure \ref{pretrain-fig}. 

\subsection{Extended Communication} 
\label{comm-extend-maddpg}
In the main text we showed that extended goal-setting in FMH lead to substantially improved performance. We also considered whether a similar approach would benefit methods which do not treat communication as goals.

\begin{figure}[ht]
%\vskip 0.2in
  \begin{minipage}[c]{0.32\textwidth}%\begin{center}
\centerline{\includegraphics[width=\columnwidth]{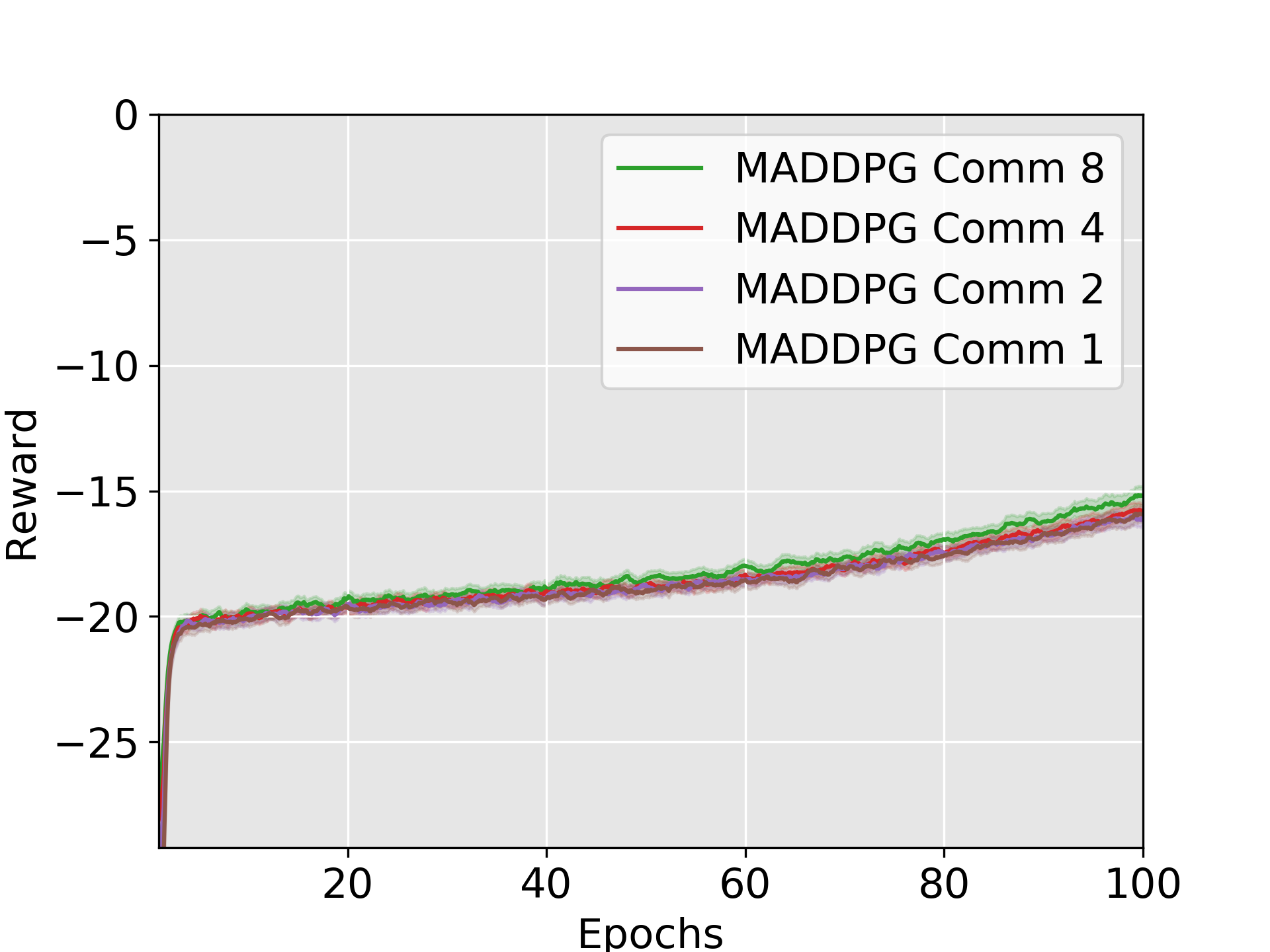}}
\end{minipage}% 
%\hfill
 \begin{minipage}[c]{0.1533\textwidth}
\caption{\\Extended communication does not significantly improve the performance of MADDPG.}
\label{comm-extend-maddpg-fig}%\end{center}
 \end{minipage}
%\vskip -0.2in
\end{figure}

We found that extended communication did not help MADDPG on the same task, with learning curves being in all cases virtually identical (Figure \ref{comm-extend-maddpg-fig}). 

\subsection{Further details on Table 1}
\label{table-scaling}
Values in the table were determined using 10 random seeds in all cases, except for the one exception of MADDPG with 10 listeners and 6 landmarks, which used 3 random seeds (training time is substantially longer as we do not share parameters). The CoM agent was trained on the synthetic task with different numbers of landmarks. Performance of the trained CoM policies was then evaluated over a period of 10 epochs on the corresponding true tasks.

Convergence was determined by comparing the mean performance in the final 5 epochs with the mean performance of a sliding window 5 epochs in width (we also take the mean across random seeds). If the mean performance within the window was within 2 percent of the final performance, and remained so for all subsequent epochs, we defined this as convergence, unless the first time this happened was within the final 10 epochs. In such a case, we define the algorithm as not having converged. For assessing the exact time of convergence in the case of FMH we report values which include the 10 epochs of pretraining.

\subsection{Cooperative Communication with 3 landmarks}
\begin{figure}[ht]
%\vskip 0.2in
  \begin{minipage}[c]{0.32\textwidth}%\begin{center}
\centerline{\includegraphics[width=\columnwidth]{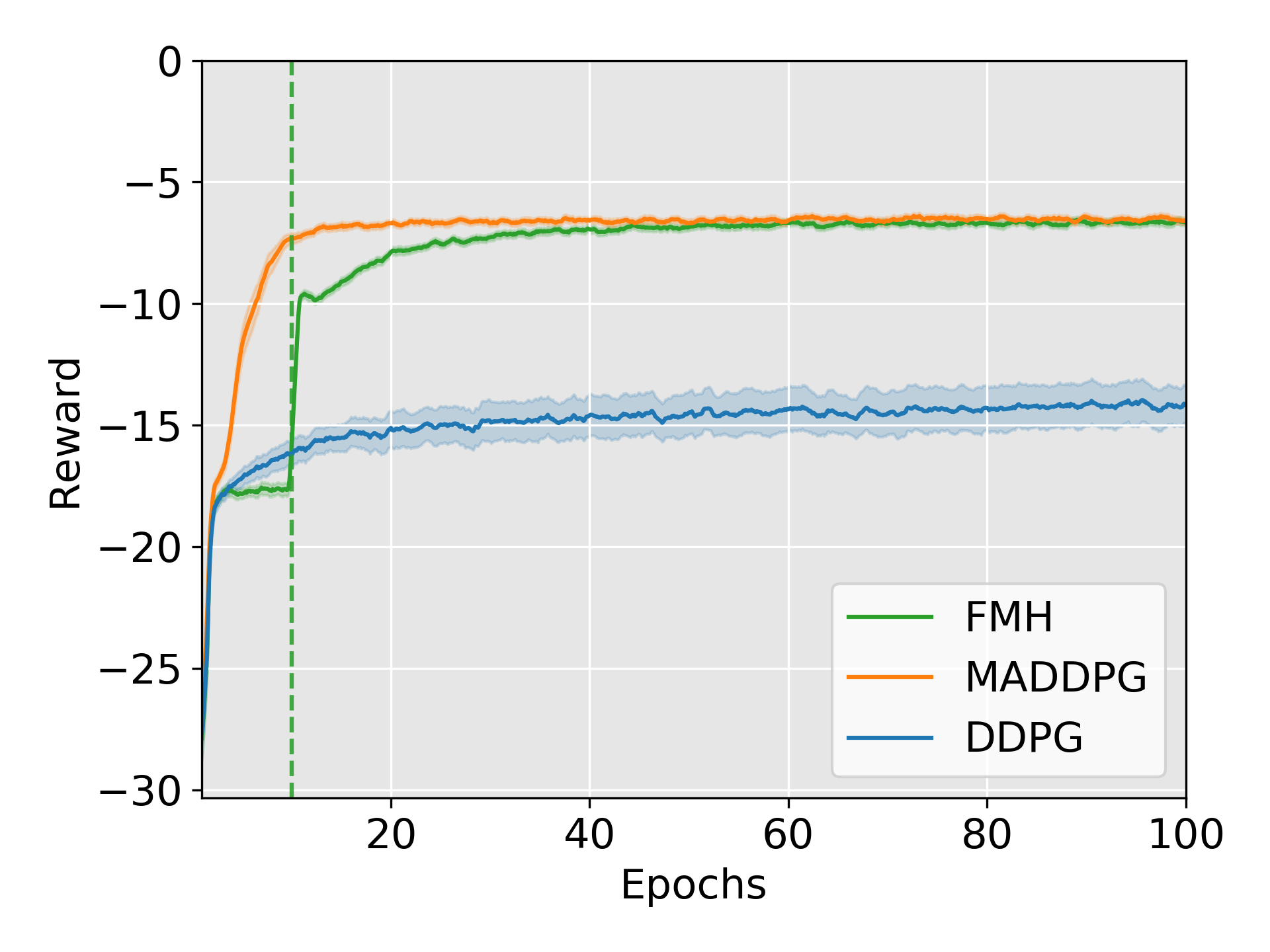}}
\end{minipage}% 
%\hfill
 \begin{minipage}[c]{0.1533\textwidth}
\caption{Cooperative Communication with 1 listener and 3 landmarks.}
\label{coop-comm-3-fig}%\end{center}
 \end{minipage}
%\vskip -0.2in
\end{figure}

For reference we show performance of the various algorithms on Cooperative Communication with 3 landmarks. Both MADDPG and FMH perform well on this task, although MADDPG reaches convergence more rapidly (Figure~\ref{coop-comm-3-fig}).

\section{Environments}
\subsection{Cooperative Communication}
\label{coop-comm-details}
We provide further details on our version of Cooperative Communication (see main text for original description). In general, we keep environment details the same as Lowe et al., including the fact that the manager only has access to the target colour. However, we also scale up the number of coloured landmarks, which we do by taking the RGB values provided in the multi-agent particle environment, $[0.65,0.15,0.15], [0.15,0.65,0.15], [0.15,0.15,0.65]$, and adding 9 more by placing them on the remaining vertices of the corresponding cube and at the centre-point of four of the faces (in RGB space). The colours for the 12 landmarks are therefore:

$ [0.65,0.15,0.15], [0.15,0.65,0.15], [0.15,0.15,0.65],$
$[0.65,0.65,0.15], [0.15,0.65,0.65], [0.65,0.15,0.65],$
$[0.65,0.65,0.65], [0.15,0.15,0.15], [0.40,0.40,0.65], $
$[0.40,0.40,0.15], [0.15,0.40,0.40], [0.65,0.40,0.40]$.

The particular colour values used for the landmarks influences the performance of RL algorithms as landmarks which have similar colours are harder for the speaker to learn to distinguish.  
\subsection{Cooperative Coordination}
\label{coop-coord-details}
We provide further details on our version of Cooperative Coordination (see main text for original description). The task provides a negative reward of -1 to each agent involved in a collisions. For DDPG and MADDPG this penalty is shared across agents, whereas in FMH only the agents involved in the collision experience this penalty. 

We also evaluated performance of trained policies in Figures \ref{coop-coord}c and \ref{coop-coord}d with slight modifications to the overall task. In the case of Figure \ref{coop-coord}c, to ensure that targets were never impossible to achieve by overlapping with the immobile manager, we moved the manager off-screen. For Figure \ref{coop-coord}d we ensured that agent positions were never initialised in a way such that they would automatically collide (such cases are rare).
\subsubsection{Coordination with a mobile manager}
\label{manager-moves}

\begin{figure}[ht]
%\vskip 0.2in
  \begin{minipage}[c]{0.32\textwidth}%\begin{center}
\centerline{\includegraphics[width=\columnwidth]{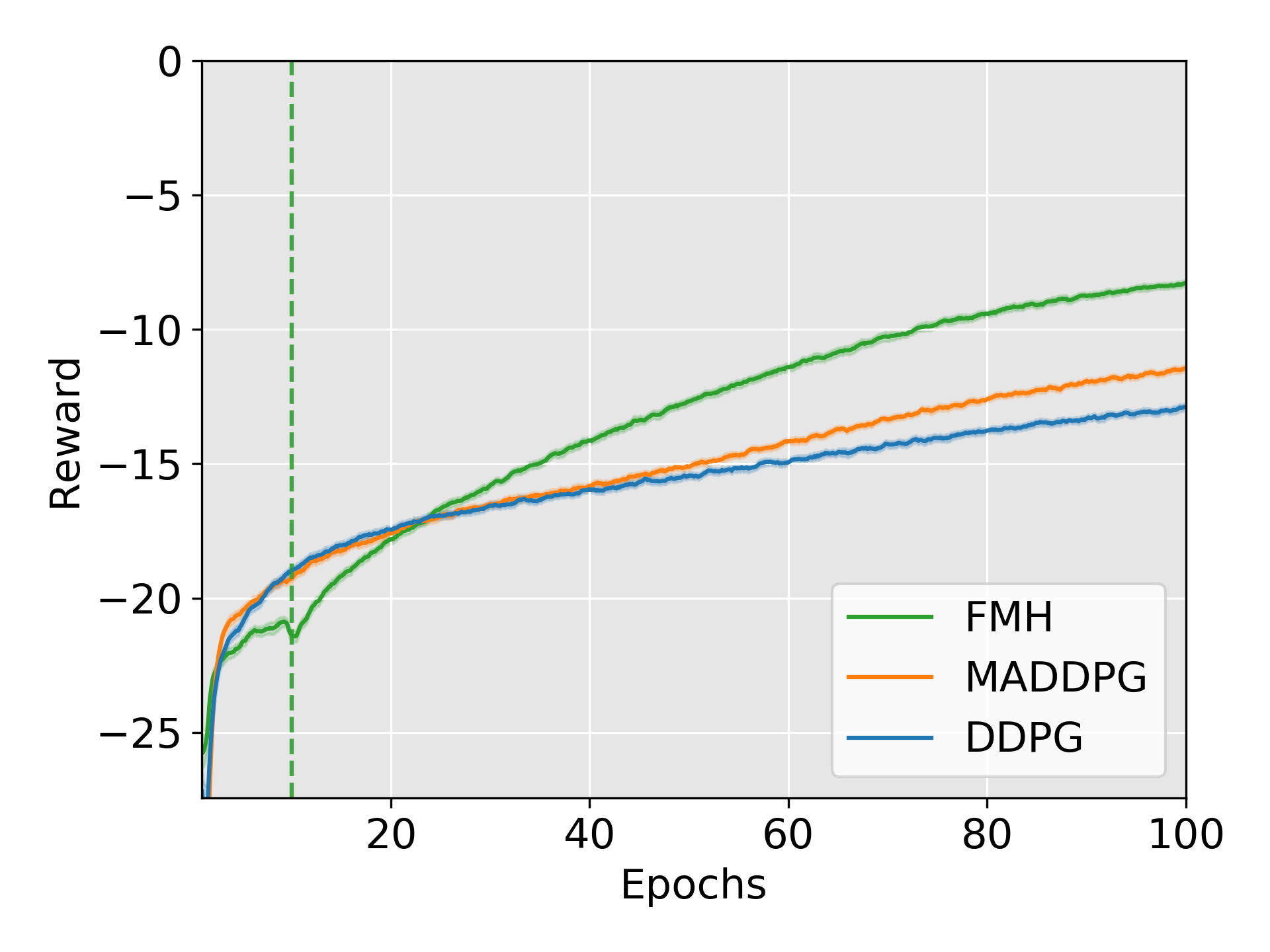}}
\end{minipage}% 
%\hfill
 \begin{minipage}[c]{0.1533\textwidth}
\caption{FMH performs well, even when the manager is required to move to cover targets whilst also setting goals for workers}
\label{manager-moves-fig}%\end{center}
 \end{minipage}
%\vskip -0.2in
\end{figure}

We implemented a version of Cooperative Coordination with 2 listeners and 1 speaker which together need to cover the 3 green targets. The manager must multi-task, directing workers to the correct targets whilst also covering these targets itself. We find that the manager learns to do this, outperforming both MADDPG and DDPG (Figure~\ref{manager-moves-fig})

\subsubsection{The Two-Near, One-Far Task}
\label{diversity}
This task is not used for training but to evaluate the performance of agents trained on a version of Cooperative Coordination in which one agent is twice as light as normal and the remaining two are twice as heavy. 

Evaluating the optimal assignments on this task can be difficult, so we assess it in the more easily interpreted TNOF environment. In TNOF, the agents start at the bottom of the environment. Two green targets are located nearby (in bottom 40 percent of screen) whereas one target is far away (in top 30 percent of screen). The x-coordinates are randomly sampled at the beginning of each episode and blue decoys are also added (one nearby, two far). 

\section{Solving a Coordination Game}
\label{coordination-game}
Some of the most popular forms of multiagent task are coordination games from microeconomics in which there are multiple good and bad Nash equilibria, and it is necessary to find the former. It is intuitively obvious that appointing one of the agents as a manager might resolve the symmetries inherent in cooperative coordination game in which agents need to take different actions to receive reward:
\begin{table}[htb!]
	\setlength{\extrarowheight}{2pt}
	\begin{tabular}{*{4}{c|}}
		\multicolumn{2}{c}{} & \multicolumn{2}{c}{Player $Y$}\\\cline{3-4}
		\multicolumn{1}{c}{} &  & $A$  & $B$ \\\cline{2-4}
		\multirow{2}*{Player $X$}  & $A$ & $(0,0)$ & $(1,1)$ \\\cline{2-4}
		& $B$ & $(1,1)$ & $(0,0)$ \\\cline{2-4}
	\end{tabular}
\end{table}

This game has two pure strategy Nash equilibria and one mixed strategy Nash equilibrium which is Pareto dominated by the pure strategies. The challenge of this game is for both agents to choose a single Pareto optimal Nash equilibrium, either $(A,B)$ or $(B,A)$.
%, either $(A,A)$ or $(B,B)$. 
\\
\\
For a matrix game, we define the feudal approach as allowing Player $X$, the manager, to specify the reward player $Y$ will receive for its actions. This is a simplification when compared to the more general setting of a Markov game in which the feudal manager can reward not only actions but also achievement of certain states.\footnote{Typically we prefer the manager to choose distal states as targets rather than actions as this requires the manager to micromanage less and so supports temporal abstraction}. In order to specify the reward, we assume that Player $X$ communicates a goal, either $g_A$ or $g_B$, prior to both players taking their actions. If Player $X$ sends $g_A$ it means that action $A$ is now rewarded for Player $Y$ and action $B$ is not. Player $X$'s rewards are unchanged, and so together this induces the following matrix game:

\begin{table}[htb]
	\setlength{\extrarowheight}{2pt}
	\begin{tabular}{*{4}{c|}}
		\multicolumn{2}{c}{} & \multicolumn{2}{c}{Player $Y$}\\\cline{3-4}
		\multicolumn{1}{c}{} &  & $A$  & $B$ \\\cline{2-4}
		\multirow{2}*{Player $X$}  & $A$ & $(0,1)$ & $(1,0)$ \\\cline{2-4}
		& $B$ & $(1,1)$ & $(0,0)$ \\\cline{2-4}
	\end{tabular}
\end{table}
Action $A$ for player $Y$ is now strictly dominant and so a rational Player $Y$ will always choose it. By iterated elimination of strictly dominated strategies we therefore find the resulting matrix game:

\begin{table}[htb]
	\setlength{\extrarowheight}{2pt}
	\begin{tabular}{*{3}{c|}}
		\multicolumn{2}{c}{} & \multicolumn{0}{c}{Player Y} \\\cline{3-3}
		\multicolumn{1}{c}{} &  & $A$   \\\cline{2-3}
		\multirow{2}*{Player $X$}  & $A$ & $(0,1)$  \\\cline{2-3}
		& $B$ & $(1,1)$  \\\cline{2-3}
	\end{tabular}
\end{table}

And so a rational Player $X$ will always choose $B$, resulting in an overall strategy of $(B,A)$ conditioned on an initial communication of $g_A$. By symmetry, we can see that conditioned on $g_B$, rational players $X$ and $Y$ will play $(A,B)$. The feudal approach therefore allows the manager to flexibly coordinate the pair of agents to either Nash equilibrium. For games involving N players, coordination can be achieved by the manager sending out N-1 goals.

\end{document}